\pdfoutput=1
\documentclass[10pt,twocolumn,letterpaper]{article}

\usepackage{iccv}
\usepackage{times}
\usepackage{epsfig}
\usepackage{graphicx}
\usepackage{amsmath}
\usepackage{amssymb}

\usepackage[breaklinks=true,bookmarks=false]{hyperref}

\iccvfinalcopy % *** Uncomment this line for the final submission

\def\iccvPaperID{****} % *** Enter the ICCV Paper ID here

\ificcvfinal\fi

\usepackage{xcolor}

\begin{document}

%%%%%%%%% TITLE
\title{ Coupled Video Frame Interpolation and Encoding with Hybrid Event Cameras for Low-Power High-Framerate Video }

\author{
Hidekazu Takahashi${}^1$\and Takefumi Nagumo${}^1$\and Kensei Jo${}^1$\and
Aumiller Andreas${}^{1,2}$\and Saeed Rad${}^2$\and Rodrigo Caye Daudt${}^2$\and
Yoshitaka Miyatani${}^1$\and Hayato Wakabayashi${}^1$\and Christian Brandli${}^2$\\
${}^1$Sony Semiconductor Solutions Corporation, Atsugi, Japan\\
${}^2$Sony Advanced Visual Sensing AG, Schlieren, Switzerland\\
{\tt\small Hidekazu.Takahashi@sony.com}
}

\maketitle
\ificcvfinal\thispagestyle{empty}\fi

%%%%%%%%% ABSTRACT
\begin{abstract}
   % Context
   Every generation of mobile devices strives to capture video at higher resolution and frame rate than previous ones.
   This quality increase also requires additional power and computation to capture and encode high-quality media.
   % Central question
   We propose a method to reduce the overall power consumption for capturing high-quality videos in mobile devices.
   % What's already known
   Using video frame interpolation (VFI), sensors can be driven at lower frame rate, which reduces sensor power consumption.
   With modern RGB hybrid event-based vision sensors (EVS), event data can be used to guide the interpolation, leading to results of much higher quality.
   % Reason for paper/research
   If applied naively, interpolation methods can be expensive and lead to large amounts of intermediate data before video is encoded.
   % Methods
   This paper proposes a video encoder that generates a bitstream for high frame rate video without explicit interpolation. 
   % The proposed method merges the motion compensation of the conventional video encoder with the motion compensation, which is the core processing of frame interpolation. 
   The proposed method estimates encoded video data (notably motion vectors) rather than frames.
   Thus, an encoded video file can be produced directly without explicitly producing intermediate frames.
   % Findings
   % Fill in if additional experiments are added
   % Significance
   % We show that the proposed method can significantly reduce the amount of power of mobile video capturing systems without noticeable effects on image quality.
\end{abstract}

%%%%%%%%% BODY TEXT
\section{Introduction}\label{sec:intro}

% Context
The latest 5G communication improves network traffic by acquiring higher throughput and low latency capabilities.
These are expected to bring new customer experiences; however, new devices and applications supported by 5G cause new challenges~\cite{khanSurveyMobileEdge2022}.
Regarding mobile video streaming, the improved bandwidth and latency have allowed us to handle videos with higher resolution.
On the other side, high-end sensors that record high-quality videos play important roles in capturing videos; their quality causes an increase of power and computation.
These can lead to overheating of the chip, and limited recording time.

% What's already known (VFI)
Video frame interpolation (VFI) addresses the problems as described previously by generating intermediate frames between keyframes.
It enables the generation of high frame rate videos while driving sensors at lower frame rate, thus reducing sensor power consumption.
Warping-based approaches are currently the main direction of VFI research, where intermediate frames are reconstructed by computing motion vectors between keyframes and sampled from them correspondingly~\cite{jiangSuperSloMoHigh2018,niklausSoftmaxSplattingVideo2020}.

% What's already known (EVS, EVS-based VFI)
Additional data between frames is often useful for performing VFI. Event-based vision sensors (EVS)~\cite{niwa297mmPitchEventBasedVision2023}, for instance, are bio-inspired sensors which record brightness changes (``events") asynchronously.
Events are recorded with high temporal resolution and can capture motion information between keyframes.
For this reason, event-assisted VFI is an active field of research~\cite{tulyakovTimeLensEventBased2021,tulyakovTimeLensEventBased2022,heTimeReplayerUnlockingPotential2022,wuVideoInterpolationEventdriven2022,kimEventBasedVideoFrame2023}.
In particular, RGB hybrid EVS~\cite{kodama122mm356MpixelRGB2023} acquire intensity images and events simultaneously and coregistered, making it an excellent match for event-based VFI and for reducing sensor power consumption.

% Reason for paper/our research
Although VFI is a good approach to reduce power consumption and computation in mobile devices by allowing us to drive the sensor readout with lower frame rates,
the interpolation method itself can be expensive and lead to large amount of intermediate data before video encoding.
In particular, naively relying on VFI would still generate a large amount of video data to be encoded, which is also a computationally expensive process.

% Central question and methods
To solve this problem, this paper proposes a video encoder that generates a bitstream for high frame rate video directly without explicitly producing intermediate frames.
The proposed encoder is designed to combine predictive coding and transformation, following modern video coding standards~\cite{sullivanOverviewHighEfficiency2012,brossOverviewVersatileVideo2021}.
Intermediate frames are coded using inter prediction and packed to a bitstream, which includes motion information and no residual signal.
They are only reconstructed by motion compensation in the decoder side during playback instead of before video encoding,
significantly reducing the resources necessary for video recording.
In addition, operations of encoding intermediate frames are lighter than those of standard video coding,
which should minimize encoding power and computation.
This paper shows a video encoder that generates a bitstream for higher frame rate video than the sensor, producing a video file directly without explicitly computing intermediate frames.

% % Contribution including central question, methods and findings
% The main contributions of the paper are as follows:
% \begin{itemize}
%     \item We propose a video encoder that generates a bitstream for higher frame rate video than the sensor, producing a video file directly without explicitly computing intermediate frames.
%     \item We show that the proposed method achieves sufficient image quality without noticeable effects on two video sequences and simulated event data.
% \end{itemize}
\section{Related Work}\label{sec:related}

\subsection{Video Coding Standards}

Video coding standards such as HEVC~\cite{sullivanOverviewHighEfficiency2012} and VVC~\cite{brossOverviewVersatileVideo2021} allow us to store video data with minimal perceptual artifacts at very high compression rates.
Such standards have evolved over time following the evolution of the hardware available for encoding and decoding data and our understanding of the human visual system.
They are still a strong competitor for learning-based video coding.

\subsection{Learning-based video coding}

There is a long history to update coding algorithms while maintaining video quality.
Learning-based video coding is one of the methods that has achieved better video quality over standard video codings~\cite{hoangRecentTrendingLearning2021}.
The neural-integrated approach~\cite{yanConvolutionalNeuralNetworkBased2018,zhaoEnhancedBiPredictionConvolutional2019,benjakEnhancedMachineLearningbased2021} is the method in which the conventional codings and the learning-based modules cooperate.
It replaces a part of modules such as prediction and loop filtering with learning-based operations.
On the other side, the neural-based~\cite{luDVCEndEndDeep2019,agustssonScaleSpaceFlowEndEnd2020} approach fully replaces the conventionally designed modules with learning-based models.
~\cite{luDVCEndEndDeep2019} is known as a base research on this region, which estimates motion vectors and residual signals, and encode them using neural networks.
It uses a pre-trained network inferring motion vectors, which is related to our work.

\subsection{Video Frame Interpolation}

Frame-based approaches~\cite{jiangSuperSloMoHigh2018,niklausSoftmaxSplattingVideo2020} rely on input from a conventional camera.
In contrast to frame-based approaches, event-based approaches~\cite{tulyakovTimeLensEventBased2021,tulyakovTimeLensEventBased2022,heTimeReplayerUnlockingPotential2022,wuVideoInterpolationEventdriven2022,kimEventBasedVideoFrame2023} rely on input from an event camera instead of a conventional camera.
An event camera can drive faster than a conventional one and capture detailed features between frames, which is well-suited to VFI.
Warping-based approaches fuse motion vector estimation with image warping to generate intermediate frames between keyframes, which is currently the main direction of VFI research.
These approaches construct the frames by motion compensation, which is related to our work.
\section{Proposed Method}\label{sec:proposed}

\begin{figure}[t]
    \centering
    \includegraphics[width=1.0\linewidth,bb=0 0 1102 1004]{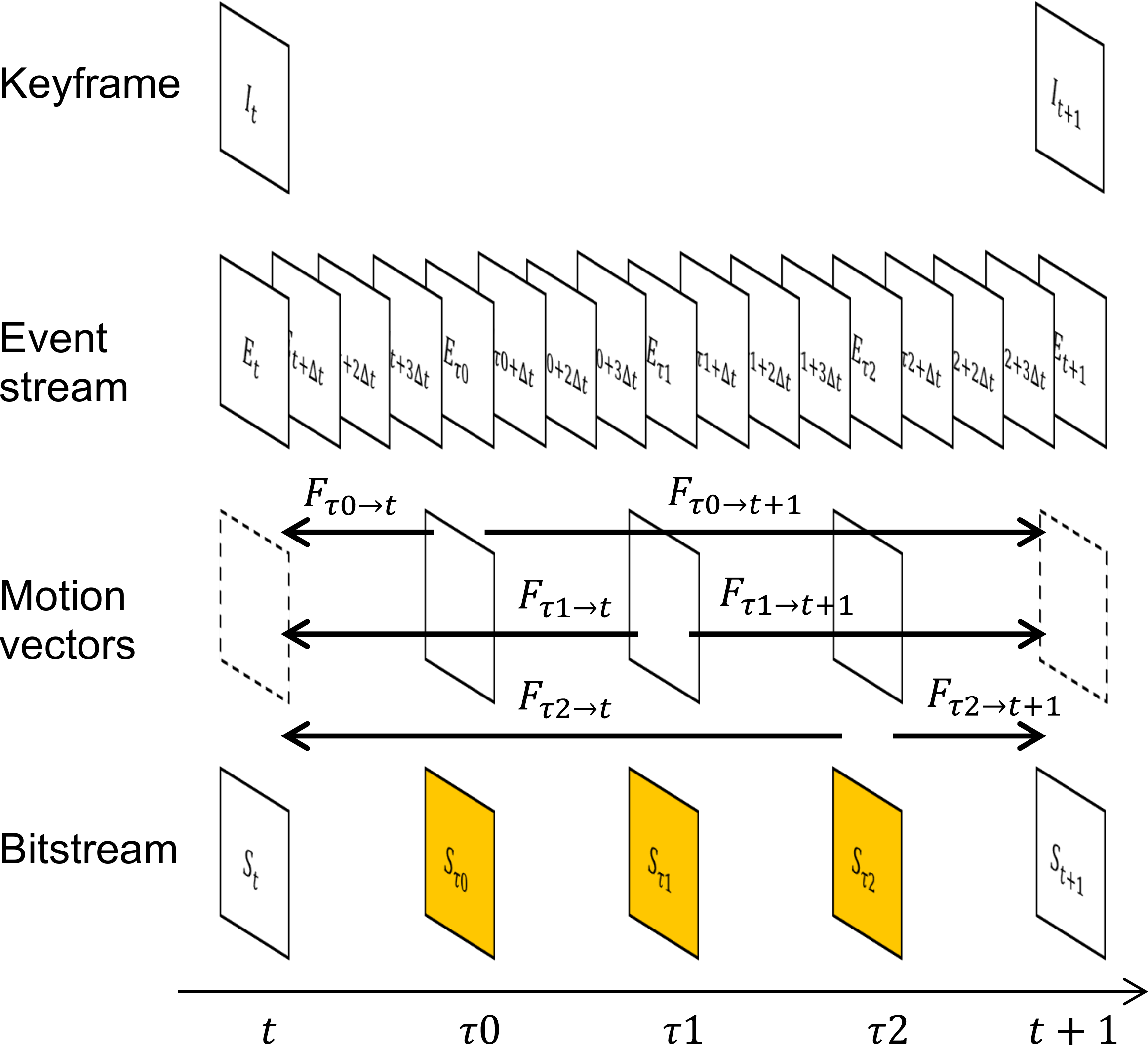}
    \caption{Example of problem formulation. 
    Three intermediate frames are generated from two keyframes and an event stream. 
    Each intermediate frame is encoded by only motion information.}
    \label{fig:problem_formula}
\end{figure}

\begin{figure*}[t]
    \centering
    \includegraphics[width=1.0\linewidth,bb=0 0 2459 1621]{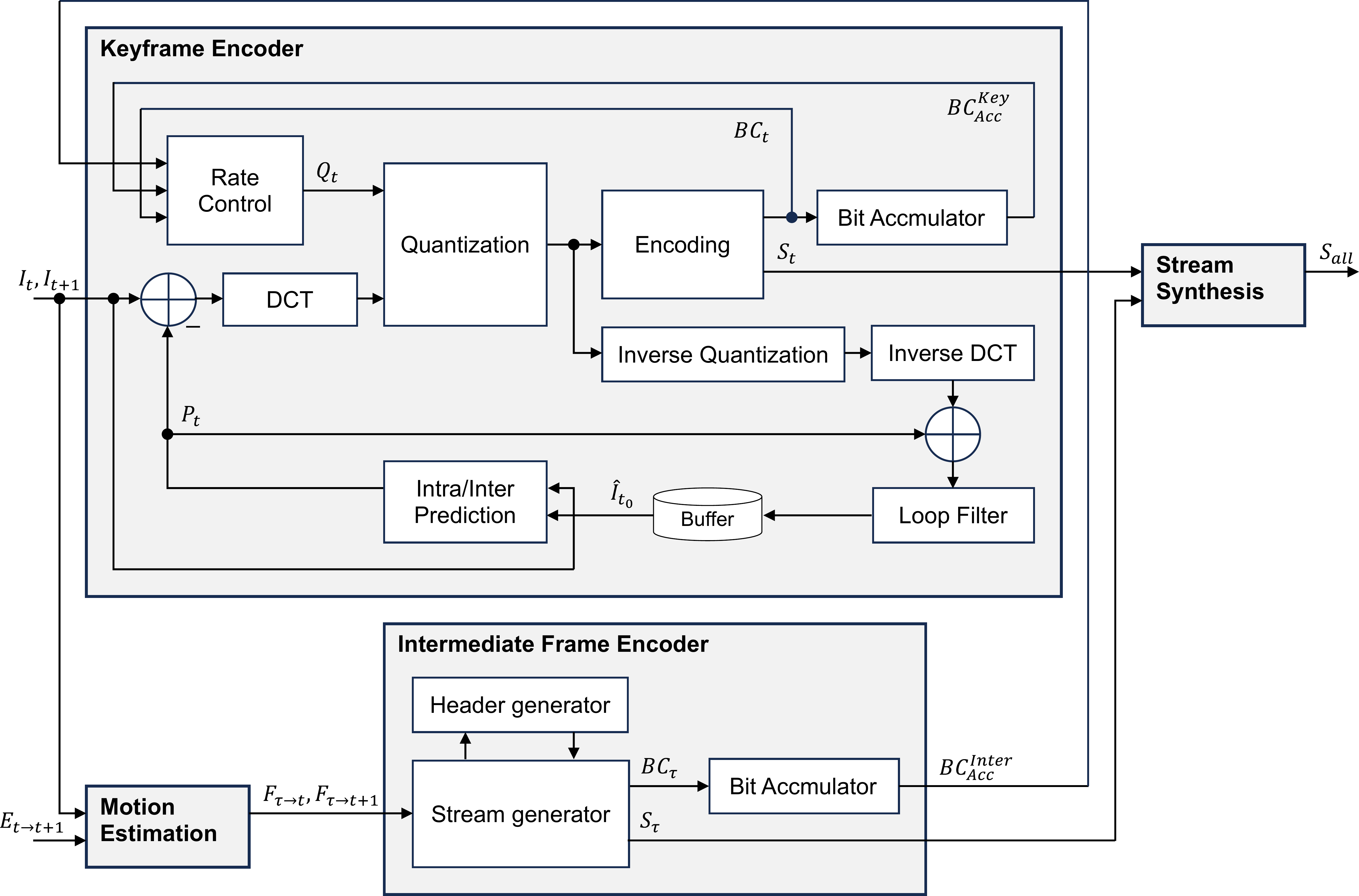}
    \caption{Overview of the proposed system.
    Key and intermediate frames are encoded separately and synthesized to generate a final bitstream.}
    \label{fig:system_overview}
\end{figure*}

\subsection{Preliminaries}

The proposed method is expanded based on video coding standards such as HEVC~\cite{sullivanOverviewHighEfficiency2012} and VVC~\cite{brossOverviewVersatileVideo2021}.
An encoding algorithm would typically proceed as follows.
A common way the encoding algorithm produces a bit stream is as follows:
The image frame is split into a block of units with a tree structure, each of which encode the operation - the tree root represents the Coding Tree Unit (CTU).
It consists of multiple Coding Units at different coarsity levels (CUs), which can be further partitioned into Prediction Units (PUs) and Transform Units (TUs).

The first picture of a video sequence is coded using intra prediction only, which does so by referring to adjacent blocks in the same frame.
Subsequent images of the movie stream are typically coded using inter prediction. This process reconstructs the current frame from a reference frame by using motion data. A inter prediction signal is generated by the encoder, applying motion compensation (warping) using motion vectors computed by the motion estimation block.
To compensate the potential error from non-ideal motion vectors, a residual signal is computed alongside latter by transforming the difference between the original block and its prediction using discrete cosine transform (DCT).
The residual signal is quantized and forwarded to the encoder, while at the same time it is locally de-quantized and inverse transformed to re-produce an approximation of the residual signal.
This is done to fix potential artifacts induced due to block-wise operations.
It may be processed by additional loop filters and the resulting filtered signal is stored in a buffer for re-use in subsequent image frames.

Each image is assigned a different coding type, called I-, P-, or B-frame.
The I (intra) coding type refers to intra prediction only, while P (predict) and B (bidirectional) represent frames coded using inter prediction with either one (forward) or two motion-compensated prediction signals, respectively.
Coding types for each image follow group of pictures (GOP), a sequence of video frames that starts with an I-frame followed by P-frames and B-frames.

\subsection{Problem Formulation}

Given a pair of consecutive keyframes $I_t$ and $I_{t+1}$ and events $E_{t\to t+1}$, we aim to generate bitstreams $S_t$, $S_{t+1}$ for keyframes $I_t$, $I_{t+1}$, and a bitstream $S_\tau$ for intermediate frames $I_\tau$ at timestamps $\tau$ ($t<\tau<t+1$).
The bitstream $S_\tau$ is generated using motion vectors $F_{\tau\to t}$ and $F_{\tau\to t+1}$.
The bitstreams $S_t$, $S_{t+1}$, and $S_\tau$ are synthesized and a final bitstream $S_{all}$ is generated.

\subsection{System Overview}

Figure~\ref{fig:system_overview} shows the overall structure of the proposed system, which consists of a keyframe encoder, optical flow estimation, intermediate frame encoder, and stream synthesis. 

The \textbf{Keyframe Encoder} block generates the bitstream $S_t$ which is given as:
\begin{equation}\label{eq:key_overview}
    S_t=\textrm{Enc}(\textrm{Quant}(\textrm{DCT}(I_t-P_t);Q_t)).
\end{equation}
where Enc, Quant, and DCT are functions of encoding, quantization, and DCT, and $P_t$ and $Q_t$ denote the predictive signal and the quantization value, respectively.
The predictive signal $P_t$ is given as:
\begin{equation}\label{eq:key_pred}
    P_t=\textrm{Pred}(I_t,\hat{I}_t).
\end{equation}
where Pred is the prediction function, and $\hat{I}_t$ is the past frame signal kept in the frame buffer.
The quantization value $Q_t$ is given as:
\begin{equation}\label{eq:key_ratectrl}
    Q_t=\textrm{RateCtrl}(BC_t,BC_{Acc}^{Key},BC_{Acc}^{Inter}).
\end{equation}
where RateCtrl is the function to control $BC_t$, $BC_{Acc}^{Key}$, and $BC_{Acc}^{Inter}$
which denote the bits spent on the current CU, those over the keyframes and the intermediate frames inside the current GOP, respectively.

The \textbf{Motion Estimation} block estimates motion vectors $F_{\tau\to t}$ and $F_{\tau\to t+1}$ which are given as:
\begin{equation}
    F_{\tau\to t}, F_{\tau\to t+1}=\textrm{ME}(I_t, E_{t\to t+1}, I_{t+1})
\end{equation}
where ME is the function of motion estimation.
You can use any methods such as proposed in ~\cite{gallegoEventbasedVisionSurvey2020}.

The \textbf{Intermediate Frame Encoder} block generates the bitstream $S_\tau$ which are give as:
\begin{equation}
    S_\tau=\textrm{SG}(F_{\tau\to t}, F_{\tau\to t+1}).
\end{equation}
where SG is the function of stream generation.
The major difference to existing encoders (including Keyframe Encoder) is that the bitstream $S_\tau$ is generated without referring to frames captured by sensors.
The encoder records the bits $BC_{Acc}^{Inter}$, and they are fed into Keyframe Encoder to decide quantization values to control bit rate.

\begin{figure}[t]
    \centering
    \includegraphics[width=0.9\linewidth,bb=0 0 857 308]{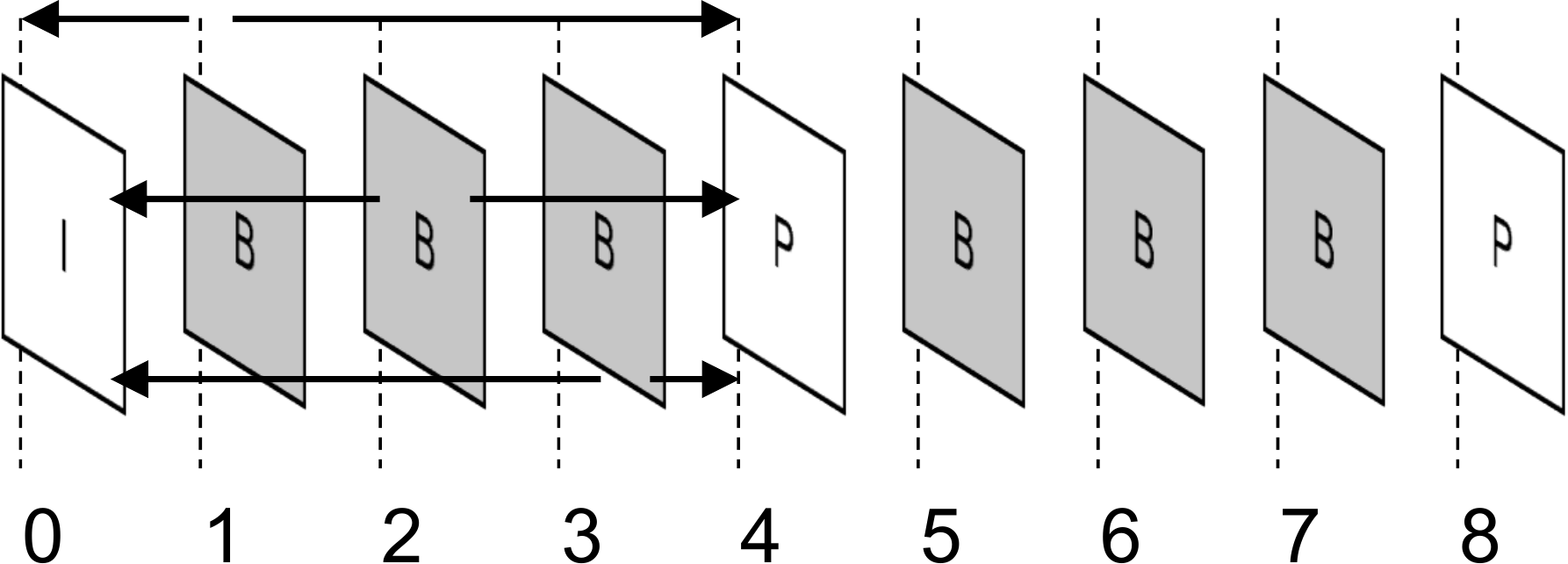}
    \caption{Example of picture structure.
    Key and intermediate frames are assigned I- or P- and B-, respectively.
    B-frames refer to keyframes, which allow B-frames to be reconstructed using image warping from keyframes.}
    \label{fig:picture_struct}
\end{figure}
The \textbf{Stream Synthesis} block synthesizes the bitstreams $S_t$ and $S_\tau$, and generates a final bitstream $S_{all}$.
The bitstream $S_{all}$ is the sequence of the bitstreams $S_t$ and $S_\tau$, which is reordered depending on the picture structure.
For instance, if the sequence is IBBBPBBBP and reference directions for P and B pictures follow Fig.~\ref{fig:picture_struct},
the decoding order should be IPBBBPBBB because P pictures should be decoded before decoding B pictures.
\section{Conclusion}\label{sec:conclude}
We propose a video encoder that generates a bitstream for high frame rate video without explicit interpolation.
The proposed method handles motion vectors between keyframes and packs them into a bitstream for higher frame rate video.
Intermediate frames are only reconstructed by motion compensation in the decoder side during playback instead of before video encoding,
significantly reducing the resources necessary for video recording.
In addition, operations of encoding intermediate frames are lighter than those of standard video coding,
which should minimize encoding power and computation.
% Our evaluation shows a limitation regarding bit rate and image quality adjustment.
% In addition, we discuss the relationship between them while analyzing encoding statistics for intermediate frames.
% While this paper has focused on the evaluation using simulation data, we will evaluate ours on real data captured by RGB hybrid EVS, which is our future work.

{\small
\bibliographystyle{ieee_fullname}
\bibliography{peregrine}
}

\end{document}